\documentclass[conference]{IEEEtran}
\usepackage{balance}
\IEEEoverridecommandlockouts
\usepackage[T1]{fontenc}
\usepackage[english]{babel}
\usepackage{graphicx}
\usepackage{xspace}
\usepackage{amsthm}
\usepackage{amsfonts}
\usepackage{amsmath}
\usepackage{booktabs}
\usepackage[capitalize]{cleveref}
\usepackage{glossaries}
\usepackage{tikz}
\usepackage{cite}
\usepackage{multirow}
\usepackage{siunitx}

\glsdisablehyper
\AtBeginDocument{\providecommand\BibTeX{{Bib\TeX}}}
\pgfkeys{/pgf/number format/.cd, fixed, precision=2}
\def\BibTeX{{\rm B\kern-.05em{\sc i\kern-.025em b}\kern-.08em
    T\kern-.1667em\lower.7ex\hbox{E}\kern-.125emX}}

\makeatletter
\newcommand\notsotiny{\@setfontsize\notsotiny\@vipt\@viipt}
\newcommand*{\toroman}[1]{\expandafter\@slowromancap\romannumeral #1@}
\makeatother
\newcommand\mperiod[1][\rlap]{#1{\;.}}
\newcommand\mcomma[1][\rlap]{#1{\;,}}

\definecolor{red}    {HTML}{b7211f}
\definecolor{orange} {HTML}{FFA500}
\definecolor{blue}   {HTML}{4169E3}
\definecolor{green}  {HTML}{147546}
\definecolor{purple} {HTML}{92268F}

\usetikzlibrary{shapes.misc}
\usetikzlibrary{arrows.meta}
\usetikzlibrary{calc}
\usetikzlibrary{fit}
\usetikzlibrary{datavisualization}
\usetikzlibrary{datavisualization.formats.functions}
\usetikzlibrary{math}
\makeatletter
\tikzset{
    >=stealth,
    font=\sffamily,
    ultra thin/.style= {line width=0.1pt},
    very thin/.style=  {line width=0.2pt},
    thin/.style=       {line width=0.4pt},%
    semithick/.style=  {line width=0.6pt},
    thick/.style=      {line width=0.8pt},
    very thick/.style= {line width=1.2pt},
    ultra thick/.style={line width=1.6pt},
    fblue/.style={fill=blue!50},
    fred/.style={fill=red!50},
    forange/.style={fill=orange!50},
    fgreen/.style={fill=green!50},
    fgray/.style={fill=gray!50},
    lblue/.style={draw=blue},
    lred/.style={draw=red},
    lorange/.style={draw=orange},
    lgreen/.style={draw=green},
    lgray/.style={draw=gray},
    AN/.style={anchor=north},
    ANW/.style={anchor=north west},
    ANE/.style={anchor=north east},
    AE/.style={anchor=east},
    AW/.style={anchor=west},
    AS/.style={anchor=south},
    ASW/.style={anchor=south west},
    ASE/.style={anchor=south east},
    AC/.style={anchor=center},
	opaque node/.code 2 args={\tikzset{opacity=#1, text opacity=#2}},
	double color fill/.code 2 args={%
		\pgfdeclareverticalshading[%
		tikz@axis@top,tikz@axis@middle,tikz@axis@bottom%
		]{diagonalfill}{100bp}{%
			color(0bp)=(tikz@axis@bottom);%
			color(50bp)=(tikz@axis@bottom);%
			color(50bp)=(tikz@axis@middle);%
			color(50bp)=(tikz@axis@top);%
			color(100bp)=(tikz@axis@top)%
		}%
		\tikzset{%
			shade,%
			left color=#1,%
			right color=#2,%
			shading=diagonalfill%
		}%
	},%
    trace0/.style = {thick, blue, solid},
    trace1/.style = {thick, red, densely dashdotdotted},
    trace2/.style = {thick, green, dashed},
    guides/.style = {gray, loosely dashed},
    fitting node/.style={
        inner sep=0pt,
        fill=none,
        draw=none,
        reset transform,
        fit={(\pgf@pathminx,\pgf@pathminy) (\pgf@pathmaxx,\pgf@pathmaxy)}
    },
    reset transform/.code={
        \pgftransformreset
    }
}
\makeatother
\newacronym{ev}{EV}{Electric Vehicle}
\newacronym{lion}{Li-ion}{Lithium-ion}
\newacronym{soc}{SoC}{State of Charge}
\newacronym{soh}{SoH}{State of Health}
\newacronym{pwm}{PWM}{Pulse Width Modulation}
\newacronym{milp}{MILP}{Mixed Integer Linear Programming}
\newacronym{bms}{BMS}{Battery Management System}
\newacronym{mosfet}{MOSFET}{Metal-Oxide-Semiconductor Field-Effect Transistor}

\newcommand{\charge}[1]{
    \ifthenelse{\equal{#1}{}}%
    {\ensuremath{Q}}%
    {\ensuremath{Q_{#1}}}%
}
\newcommand{\maxcharge}[1]{%
    \ifthenelse{\equal{#1}{}}%
    {\ensuremath{\charge{max}}}%
    {\ensuremath{\charge{max,#1}}}%
}
\newcommand{\chargeloss}[1]{%
    \ifthenelse{\equal{#1}{}}%
    {\ensuremath{\charge{loss}}}%
    {\ensuremath{\charge{loss,#1}}}%
}

\newcommand{\soh}[1]{
    \ifthenelse{\equal{#1}{}}%
    {\ensuremath{SoH}}%
    {\ensuremath{SoH_{#1}}}%
}
\newcommand{\current}[1]{
    \ifthenelse{\equal{#1}{}}%
    {\ensuremath{I}}%
    {\ensuremath{I_{#1}}}%
}
\newcommand{\selfdischarge}[1]{
    \ifthenelse{\equal{#1}{}}%
    {\ensuremath{\current{sr}}}%
    {\ensuremath{\current{sr,#1}}}%
}
\newcommand{\rate}[1]{
    \ifthenelse{\equal{#1}{}}%
    {\ensuremath{\eta}}%
    {\ensuremath{\eta_{#1}}}%
}
\newcommand{\ratecharge}[1]{
    \ifthenelse{\equal{#1}{}}%
    {\ensuremath{\rate{c}}}%
    {\ensuremath{\rate{c,#1}}}%
}
\newcommand{\ratedischarge}[1]{
    \ifthenelse{\equal{#1}{}}%
    {\ensuremath{\rate{d}}}%
    {\ensuremath{\rate{d,#1}}}%
}
\newcommand{\ahthrp}[1]{
    \ifthenelse{\equal{#1}{}}%
    {\ensuremath{Ah_{thrp}}}%
    {\ensuremath{Ah_{thrp,#1}}}%
}
\newcommand{\crate}[1]{
    \ifthenelse{\equal{#1}{}}%
    {\ensuremath{C_{rate}}}%
    {\ensuremath{C_{rate,#1}}}%
}
\newcommand{\ct}[1]{
    \ifthenelse{\equal{#1}{}}%
    {\ensuremath{CT}}%
    {\ensuremath{CT_{#1}}}%
}

\newcommand{\drseg}[2]{\ensuremath{\langle I_{#2}, t_{#1}, t_{#2} \rangle}}

\begin{document}

\bstctlcite{IEEEexample:BSTcontrol}

\setlength{\abovedisplayskip}{3pt}
\setlength{\belowdisplayskip}{3pt}
\IEEEaftertitletext{\vspace{-2.00\baselineskip}}

\title{To Balance or to Not? Battery Aging-Aware Active Cell Balancing for Electric Vehicles}

\author{
\IEEEauthorblockN{%
    Enrico Fraccaroli\textsuperscript{1, 3}, %
    Seongik Jang\textsuperscript{2}, %
    Logan Stach\textsuperscript{3}, %
    Hoeseok Yang\textsuperscript{4}, %
    Sangyoung Park\textsuperscript{5} %
    and Samarjit Chakraborty\textsuperscript{3}
}
\IEEEauthorblockA{\textsuperscript{1}\textit{University of Verona, Italy.}}%
\IEEEauthorblockA{\textsuperscript{2}\textit{Hyundai Motor Company, South Korea.}}%
\IEEEauthorblockA{\textsuperscript{3}\textit{University of North Carolina at Chapel Hill, NC, USA.}}%
\IEEEauthorblockA{\textsuperscript{4}\textit{Santa Clara University, CA, USA.}}%
\IEEEauthorblockA{\textsuperscript{5}\textit{Technical University of Berlin, Germany.}}%
\thanks{This study was supported by the NSF grant No. 2038960, and from the European Union's Horizon Europe research and innovation program under the Marie Sklodowska-Curie grant No. 101109243. A preliminary version of this paper is due to appear in VLSI Design 2024.}%
}

\maketitle

\begin{abstract}
Due to manufacturing variabilities and temperature gradients within an electric vehicle's battery pack, the capacities of cells in it decrease differently over time.
This reduces the usable capacity of the battery -- the charge levels of one or more cells might be at the minimum threshold while most of the other cells have residual charge.
Active cell balancing (i.e., transferring charge among cells) can equalize their charge levels, thereby increasing the battery pack's usable capacity.
But performing balancing means additional charge transfer, which can result in energy loss and cell aging, akin to memory aging in storage technologies due to writing.
This paper studies when cell balancing should be optimally triggered to minimize aging while maintaining the necessary driving capability. 
In particular, we propose optimization strategies for cell balancing while minimizing their impact on aging. By borrowing terminology from the storage domain, we refer to this as ``wear leveling-aware'' active balancing.
\end{abstract}


\section{Introduction}\label{sec:introduction}
The widespread adoption of \glspl{ev} faces a key challenge -- battery aging~\cite{ChakrabortyLBFCPKLA12, GeorgakosSSC13}.
How to mitigate aging has attracted considerable attention since the battery is the most expensive component of an \gls{ev}.
Aging diminishes the charge retention capacity of cells and poses a problem in any \gls{lion} battery pack~\cite{ParkZC17}. 
Because of manufacturing variability and temperature gradient within a pack, its cells can age at different rates~\cite{Baumhoefer2014Production}.
As a result, the charge level of cells in a pack can differ despite them all being subjected to the same charging and discharging current.

In the example of \cref{fig:balancing_driving}, an imbalanced battery pack is dealt with by using \emph{cell balancing}~\cite{KauerNSC17, Narayanaswamy2018Multi, NarayanaswamySL19}, highlighting two key issues:
(1)~different maximum capacities of cells ($\maxcharge{i}$) caused by manufacturing deviation or differential aging, and (2)~different total charge in each cell ($\charge{i}$) at any time.
Although in this example, every cell is charged or discharged at the same rate, because of the differences in their capacities $\maxcharge{i}$, their charge levels are different at any point in time.
One of the main consequences of such a charge imbalance is a \emph{reduction in the usable capacity of the battery pack}.
\cref{fig:balancing_driving} illustrates the problem where, after the \gls{ev} has driven for some time, the charge level of the second cell reduces to zero, while the first one still has charge left in it.
Once the charge level of a cell goes below or above certain thresholds (e.g., below $20\%$ or above $100\%$)~\cite{Ovejas2019State}, it should not be discharged or charged any further.
Violating these thresholds might not have an immediate impact on the battery itself.
Still, if repeated, it can lead to permanent changes in the electro-chemical properties of the battery and sub-optimal performances.
Two solutions can extend the usage of the battery: (1)~charging the pack or (2)~ transferring charges between cells so that all of them have some charge.
The latter is known as \emph{cell balancing}~\cite{Roy2019Multi, KauerNSLC15, KauerNLSC15}, and the objective is to equalize the \gls{soc} between the cells of a pack.
The \gls{soc} represents a cell's charge level relative to its capacity and is expressed as a percentage.
\begin{figure}[tbp]
    \centering
    \newcommand\SIZE{0.35cm}%
\newcommand\SPACE{0.2cm}%
\begin{tikzpicture}[
    every node/.append style={align=center, font=\footnotesize\sffamily, inner sep=0pt, outer sep=0},
    cell/.style = {draw, ASW, minimum width=0.425cm, rounded corners=1pt},
    charge/.style = {cell, AS, inner sep=0pt, outer sep=0, fill=green!25},
    arrow/.style  = {->, shorten >= 3pt, shorten <= 3pt},
]
\node[cell,   minimum height=3.00*\SIZE] (C1a) at (0,0) {};
\node[cell,   minimum height=2.00*\SIZE] (C2a) at ($(C1a.south east)+(1*\SPACE,0)$) {};
\node[charge, minimum height=3.00*\SIZE] (Q1a) at (C1a.south) {};
\node[charge, minimum height=2.00*\SIZE] (Q2a) at (C2a.south) {};
\node[draw, red, inner sep=3pt, rounded corners=2pt, fit=(C1a)(C2a)] (BP) {};

\node[cell,   minimum height=3.00*\SIZE] (C1b) at ($(C2a.south east)+(6*\SPACE,0)$) {};
\node[cell,   minimum height=2.00*\SIZE] (C2b) at ($(C1b.south east)+(1*\SPACE,0)$) {};
\node[charge, minimum height=1.20*\SIZE] (Q1b) at (C1b.south) {};
\node[charge, minimum height=0.20*\SIZE] (Q2b) at (C2b.south) {};

\node[cell,   minimum height=3.00*\SIZE] (C1c) at ($(C2b.south east)+(8*\SPACE,0)$) {};
\node[cell,   minimum height=2.00*\SIZE] (C2c) at ($(C1c.south east)+(1*\SPACE,0)$) {};
\node[charge, minimum height=0.7*\SIZE] (Q1c) at (C1c.south) {};
\node[charge, minimum height=0.7*\SIZE] (Q2c) at (C2c.south) {};

\node[cell,   minimum height=3.0*\SIZE] (C1d) at ($(C2c.south east)+(6*\SPACE,0)$) {};
\node[cell,   minimum height=2.0*\SIZE] (C2d) at ($(C1d.south east)+(1*\SPACE,0)$) {};
\node[charge, minimum height=0.2*\SIZE] (Q1d) at (C1d.south) {};
\node[charge, minimum height=0.2*\SIZE] (Q2d) at (C2d.south) {};

\draw[arrow] ($(C2a.south east)+(3pt,\SIZE)$) -- ($(C1b.south west)+(0,\SIZE)$)
    node[midway, above, outer sep=3pt]{\emph{drive}};
\draw[arrow] ($(C2b.south east)+(0,\SIZE)$) -- ($(C1c.south west)+(0,\SIZE)$)
    node[midway, above, outer sep=3pt]{\emph{balance}};
\draw[arrow] ($(C2c.south east)+(0,\SIZE)$) -- ($(C1d.south west)+(0,\SIZE)$)
    node[midway, above, outer sep=3pt]{\emph{drive}};
    
\node[AS, outer sep=1pt] at (BP.north) {Battery Pack};
\node[AS, outer sep=2pt] at ($(C1b.north)-(0.050cm,0)$) {\maxcharge{1}};
\node[AS, outer sep=2pt] at ($(C2b.north)+(0.2cm,0)$) {\maxcharge{2}};
\node[AS, outer sep=2pt] at (Q1b.north) {\charge{1}};
\node[AS, outer sep=2pt] at (Q2b.north) {\charge{2}};
\end{tikzpicture}%
    \caption{Example of battery usage, where \emph{cell balancing} can increase the \emph{driving range} of a typical battery pack.}
    \label{fig:balancing_driving}
    \vspace*{-1.5em}
\end{figure}
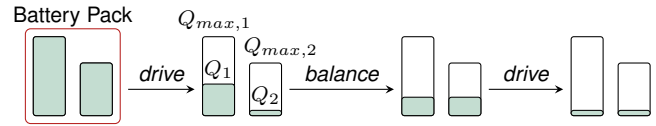\\

\noindent  
\textbf{Innovations in this paper:}
Where there is a considerable volume of work on developing optimal cell balancing strategies and architectures~\cite{NarayanaswamyPS18, SteinhorstKMNLC16, SteinhorstSCKLL16, NarayanaswamySLKC14}, active cell balancing comes at a price.
Migrating charges causes energy loss and uneven aging of cells, decreasing their capacity and increasing charge imbalance.
We should use it cautiously and do it on an \emph{as-needed basis} to minimize its negative effect on aging instead of doing it at \emph{every} opportunity.
\emph{When} should we trigger active cell balancing, and by \emph{how much} should we balance the battery pack?
These are non-trivial questions that previous studies have not addressed yet, and this paper aims to answer them.
Towards this, we assume that the \gls{ev} has to complete a \emph{mission}, defined by the distances it travels, its times, and the amount of charging/discharging current.
Given such a mission, our goal is to determine the balancing actions to accomplish it while minimizing the aging of the cells.
Borrowing terminology from the storage technologies domain, we refer to this as \emph{wear leveling}-aware, active cell balancing.

This paper shows that determining the optimal balancing schedule is an optimization problem.
The main contributions are formulating and solving this problem, which fills an essential gap in the active cell balancing literature.
An assumption we make is the notion of a \emph{mission} and our complete knowledge about it.
We encourage the readers to see this as a foundational step towards solving the more general and realistic problem, where the ``mission'' is not entirely known, and its various components are only specified in a stochastic setting.
The resulting stochastic optimization problem can be solved by leveraging the solution we propose in this paper.

\section{Background}\label{sec:background}
Much literature exists on automotive embedded systems and software~\cite{ChakrabortyFCGW16, ChangC16}.
While the issue of active power management arises in battery-operated devices~\cite{PetersPCMPC16, GuC08a, GuC08b}, a different form of power management in combination with automotive embedded systems arises in electric vehicles.
This section introduces this and outlines basic concepts related to battery aging and some prominent cell-balancing architectures.
It then explains the non-neighbor active cell balancing architecture we rely upon.

\subsection{Battery aging models}
\label{subsec:aging}
Aging-induced degradation of \gls{lion} batteries with usage is inevitable, and both external and internal factors can influence it.
There are several approaches to model battery degradation, and in general, they fall into two categories: empirical~\cite{Wang2014Degradation} and electro-chemical~\cite{Jin2017Physically} models.
As the name implies, the former is a simple and effective model based on parameters fitted from extensive measurement data.
The latter describes and mimics the underlying electrochemical processes that occur inside a battery during its lifetime.
\cref{subsec:battery_model} provides a detailed description of the model we rely upon in this paper, specifically, the empirical model presented in~\cite{Wang2014Degradation}.

\subsection{Cell balancing architectures and strategies}
\label{subsec:balancing_architectures}
There are two prominent families of cell balancing architectures: \emph{passive} and \emph{active}.
\emph{Passive} balancing equalizes the \gls{soc} across cells by dissipating excess energy from cells to reach the charge level of the cell with the lowest \gls{soc}.
Although simple, this strategy wastes energy that could instead be used for driving.
It also increases the temperature of the pack and further accelerates aging (see~\cite{Wang2014Degradation} for more details).
\emph{Active} balancing equalizes \gls{soc} by migrating charge among cells.
It is more advantageous and has been extensively studied in the literature recently.
Follows a list of the most prominent active cell balancing architectures and strategies.

Depending on the energy storage element, we could consider several variations of the active cell balancing architectures in this work.
There are three prominent families of architectures, i.e., capacitor-based~\cite{Baughman2008Double}, inductor-based~\cite{Kauer2013Modular}, and transformer-based~\cite{Narayanaswamy2017Modular}.
In this section, we first discuss a capacitor-based one, followed by an inductor-based one, and finally, we provide a lengthy description of a transformer-based one that we relied upon in this work.
The approach in this paper is compatible with any of these variations; we need access to the software controlling \emph{when} balancing is activated.

In this paper, we rely upon a transformer-based architecture that enables exchanging charges between non-adjacent cells, improving balancing efficiency and time~\cite{Narayanaswamy2017Modular}.
The proposed balancing architecture equips each cell with a balancing module composed of the flyback transformer and several switches.
It then controls each switch, establishing a path for transferring charges from the source to the destination cell, passing through a temporary energy buffer.
Thanks to its topology, charges can be physically transmitted and received simultaneously among multiple pairs of cells.
With the term \emph{single transfer cycle}, we refer to the time it takes for the software-based controller to operate the control signals that actuate the switches and transfer energy from one cell to another.
The time it takes to perform the whole transfer operation is identified by 1) the charging of the first winding and 2) the transfer of the charges to the non-adjacent cell.
Furthermore, the transfer time depends on the pair of cells involved in the operation (more details on this in \cref{sec:balancing}).
For the remainder of the paper, we identify the time it takes to perform \emph{single transfer cycle} from the $i$-th to the $j$-th cell with $T_{c}(i, j)$.

\section{Related work}\label{sec:related_work}
There are several balancing strategies for equalizing battery charges.
\cite{Kauer2013Modular} proposes an \emph{inductor-based} balancing architecture that allows concurrent charge transfers between non-adjacent cells.
It uses a heuristic-based balancing strategy that selects a set of charge transfer pairs and the appropriate architecture that enables those transfers.
A \emph{transformer-based} balancing architecture that enables concurrent charge transfers between non-adjacent cells is studied in~\cite{Narayanaswamy2017Modular}.
Here, a hybrid balancing strategy transfers charges between individual cells and from a single cell to a group of cells and vice versa, with increased energy efficiency and low balancing time.
\cite{Roy2019Optimal} proposes an active balancing strategy focusing on minimizing energy loss and balancing time.
While these strategies might achieve their objectives, they do not consider the aspects related to batteries' \gls{soh}.
They all consider cells to be identical, disregarding the effect of manufacturing process deviation on the physical aspects of the batteries and, consequently, on balancing.
In contrast, \cite{Lamprecht2019Enhancing} considers cells with different charge/discharge rates and formulates a balancing strategy that compensates for it.
It extends the usable time when discharging and reduces the charging time through preconditioning.
However, its primary goal is to extend the battery's usable capacity, and it does not aim to minimize the impact on cell aging.

Other studies propose \gls{soh}-aware cell balancing strategies.
Recently, \cite{Kremer2021Active} described an active cell balancing strategy that extends battery pack lifespan by mitigating the thermal gradient inside the pack.
However, it considers an abstracted balancing behavior without concrete consideration of actual balancing operations.
\cite{Probstl2018Soh} proposes a heuristic-based active cell balancing.
After studying the aging model, it concludes that letting weaker cells rest can extend the battery pack's lifespan.
It proposes a balancing strategy where healthier cells help the other ones discharge less load current.
However, it assumes a simplified balancing operation instead of a realistic transferring process.
The heuristic strategy cannot be based on a quantitative optimization of charge transfer.

This paper proposes an optimized balancing strategy based on the quantitative model of the imbalance evolution over time, with an analytic model of the balancing operation.

\section{System Model}\label{sec:system}
This section shows the model of the batteries, the driving profile we use to formulate our optimal balancing strategy and our analytic model of active cell balancing architecture.

\subsection{Battery Model}
\label{subsec:battery_model}
In this paper, we consider the typical battery pack mounted on \glspl{ev}, consisting of 96 series-connected modules, and each module is composed of 24 parallel connected cells.
Let us assume that the parallel cells are electronically indistinguishable, so the charging and discharging currents are evenly distributed between parallel cells.
Ideally, each manufactured cell has a nominal capacity of \SI{2.5}{\ampere\hour}; however, it eventually differs from the nominal value due to different factors, from manufacturing process variation to aging.
The ratio of the current maximum capacity of a cell to the nominal one of a new cell is a good indicator of its degree of aging, thus called \gls{soh}, and is expressed as a percentage.

Self-discharging is a chemical phenomenon that causes batteries to lose charge even when not connected to an electrical load.
These are usually called self-discharging currents and are the main reason for charge imbalance inside battery packs~\cite{Andrea2010Battery}.
Furthermore, \gls{lion} battery cells are known to charge and discharge at different rates~\cite{Yang2018Study}, which represents another source of imbalance.
Throughout this paper, we denote with $\maxcharge{i}$ the nominal capacity of the $i$-th cell and its remaining charge with $\charge{i}$ while we denote its self-discharge current with $\selfdischarge{i}$.
Charge and discharge rates are denoted as $\ratecharge{i}$ and $\ratedischarge{i}$, respectively, and describe the efficiency by which electrons are transferred from and to the cells.

The battery degrades with repeated charge and discharge cycles.
We can compute the percentage of capacity lost during the lifetime of the $i$-th cell as follows~\cite{Wang2014Degradation}:
\begin{equation}\label{eq:capacity_lost_general}
    \chargeloss{i} = a \cdot e^{b \cdot \crate{i}} \cdot \ahthrp{i} \mcomma
\end{equation}
where $\crate{i}$ is the measured ratio of current to the maximum capacity, and $\ahthrp{i}$ is the amount of charged and discharged current.
Meanwhile, $a$ and $b$ are tunable temperature-dependent coefficients.
\cref{sec:balancing} explains how both $\crate{i}$ and $\ahthrp{i}$ are computed.
We can compute the effect of aging on the \gls{soh} for the $i$-th cell as follows:
\begin{equation}
    \soh{i} \;[\%] = 100 - \chargeloss{i} \mperiod
\end{equation}
When the \gls{soh} reaches 80\% (i.e., we have lost more than 20\%), the battery is considered to be in its end-of-life state (or second-life state)~\cite{Balagopal2015State}. Ideally, it becomes inefficient if installed on an \gls{ev}; however, these second-life batteries still have some use in non-automotive applications.

\subsection{Driving Profile Model}
\label{subsec:driving_profile}
We consider the driving profile (mission) as a finite list of $n$ segments, each representing the current usage and its length in terms of time.
The current usage can be any of three among discharging ($I > 0$), charging ($I < 0$), and idle current ($I = 0$).
We can formally describe the driving profile as a sequence of tuples as follows:
\begin{equation}
    \mathcal{M} = \langle \drseg{0}{1}, ..., \drseg{k-1}{k}, ..., \drseg{n-1}{n} \rangle \mcomma
\end{equation}
where $\current{k}$ is the current usage during the whole segment, while $t_{k-1}$ and $t_{k}$ are the starting and ending times of the $k$-th segment, respectively.
Without loss of generality, we assume that the current usage inside a segment remains constant.
Nevertheless, we can consider varying currents by splitting a segment into smaller segments with different current usages.
Given a mission, we can model the current usage at time $t$ as a piecewise constant function as follows:
\begin{equation}\label{eq:current_value}
    \current{}(t) = \current{k} \mid t_{k-1} \leq t < t_{k} \mperiod
\end{equation}
We can formulate the remaining charge amount of the $i$-th cell after the $k$-th segment as follows:
\begin{equation}\label{eq:charge}
    \charge{i}^{k} = \charge{i}^{0} - \int_{0}^{t_{k}} (\rate{i} \cdot \current{}(t) + \selfdischarge{i}) \; dt \mcomma
\end{equation}
where $\rate{i}$ is $\ratecharge{i}$ when we are charging, $\ratedischarge{i}$ when we are discharging, or $0$ when idle.

\subsection{Analytic Model of Active Cell Balancing}
\label{subsec:balancing_model}
Several active cell balancing architectures exist, as discussed in \cref{sec:background}.
This paper uses the balancing architecture proposed in~\cite{Narayanaswamy2017Modular}, which enables concurrent charge transfers between non-neighbor cells within the maximum distance $d$.
The value $d$ is computed based on the voltage the switches composing the architecture can withstand.
Although concurrent charge transfer is possible, balancing paths cannot overlap between multiple pairs as it would cause a short circuit.
The following equations are \emph{inspired} by the work in~\cite{Roy2019Optimal}, and we \emph{refine} them to enable the novel formulation of \cref{sec:balancing}.

The transmitted charge from the $i$-th cell to another one during a single transfer operation is computed as follows:
\begin{equation}\label{eq:tx}
\resizebox{0.9\columnwidth}{!}{$
    \charge{tx}(i) = 
        \ln \left(\frac{V_{i}}{V_{i} - \current{peak} \cdot R_s(i)}\right) \cdot
        \frac{L \cdot V_{i}}{R_s(i)^2} -
        \frac{L \cdot \current{peak}}{R_s(i)} \mcomma
$}
\end{equation}
where $V_{i}$ is the voltage of the $i$-th cell, $R_s(i)$ is the parasitic resistances of the circuit around it, $L$ is the inductance of the secondary winding of the transformer, and $\current{peak}$ denotes the peak current that flown through it.
Specifically, $R_s(i)$ considers the source cell resistance, the parasitic resistances of its winding, and the parasitic resistances of its switches.

The amount of charge that the $i$-th cell receives from the $j$-th one in a single transfer operation is computed as follows:
\begin{equation}\label{eq:rx}
\resizebox{0.9\columnwidth}{!}{$
    \charge{rx}(i, j) =%
        \ln \left(\frac{V_{i}}{V_{i} + \current{peak} \cdot R_d(i, j)}\right) \cdot%
        \frac{L \cdot V_{i}}{R_d(i, j)^2} +%
        \frac{L \cdot \current{peak}}{R_d(i, j)} \mcomma%
$}
\end{equation}
where $R_d(i, j)$ is the total resistance in the path from the receiving cell $i$ to the source cell $j$ and is directly proportional to the distance between them.
Specifically, $R_d(i, j)$ considers the source cell resistance, the parasitic resistances of its winding, and all the parasitic resistances of the switches encountered on the path.
It does not include the destination cell's resistance, which is already considered by $R_s(i)$.

\section{Balancing methodology}\label{sec:balancing}
This section proposes our balancing strategy accounting for different cell capacities, called \emph{wear leveling-aware} active cell balancing.
This strategy aims to minimize unnecessary balancing operations based upon a limited knowledge of our future missions, i.e., only for a given time window.
Conversely, the state-of-the-art balancing strategy proposed in~\cite{Roy2019Optimal}, which we call \emph{opportunistic} balancing, assumes to equalize the charge of cells at every opportunity, regardless of future missions.

The baseline for comparing balancing strategies is the \emph{no balancing} approach, which has no impact on the aging of the cells.
While \emph{opportunistic} balancing instead exploits every single \emph{idle} period to balance the \gls{soc} of the cells.
It effectively keeps the charge levels equalized at all times but leads to accelerated aging of the cells.
However, \emph{wear leveling-aware} balancing delays the balancing operation and performs it only when necessary, i.e., when the \gls{soc} of some cells is about to go below 20\% during the next drive activity.
We can also set a higher threshold to increase our safety margin.

To summarize the effect of each strategy on aging.
The baseline is the \emph{no-balancing} strategy, which has a low impact on aging, but we might not be able to complete our mission with it.
\emph{Wear leveling-aware} balancing speeds up aging and ensures we can complete future missions.
Similarly, \emph{opportunistic} ensures we can complete future missions at the expense of faster aging.
This paper wants to find the trade-off between the two opposing strategies.

\subsection{Wear leveling-aware balancing}
\label{subsec:wla_balancing}
In this paper, we propose an optimal way to perform balancing that guarantees fulfilling the foreseen future missions when we know the details of the next day's mission.
The idea is to receive the details of the next day's mission during the night, i.e., when the vehicle is inactive and probably charging.

We have already established that we cannot further discharge serially connected cells when even one of them runs out of charge.
The \emph{usable capacity}, namely $\charge{UC}$, can be defined as the minimum remaining charge between cells.
When we know the mission for the following day, we can quantitatively calculate the remaining charge of each cell after each discharge activity by using \cref{eq:charge}.
Due to the different self-discharge currents and nominal capacities, the imbalance may get exacerbated, making the following activities unachievable without balancing.
In some cases, we would need to perform balancing during idle segments.
The set of indices of idle segments up to the $k$-th segment can be defined as follows:
\begin{equation}
    \mathcal{L}_{k} = \{l|  1 \leq l \leq k, \current{l} = 0\} \mperiod
\end{equation}

While the voltage levels of cells are assumed to remain constant during a single balancing event, they may vary between balancing events as the charge level may vary severely according to usage.
The analytic models of transmitted and received charges shown in \cref{eq:tx,,eq:rx} must be extended to account for different voltage levels at different idle segments.
As such, we need to change them as follows:
\begin{equation}
\resizebox{0.875\columnwidth}{!}{$
\charge{tx}^{l}(i)~ =
    \ln \left(\frac{V_{i}^{l}}{V_{i}^{l} - \current{peak} \cdot R_s(i)}\right) \cdot
    \frac{L \cdot V_{i}^{l}}{R_s(i)^2} \text{-}
    \frac{L \cdot \current{peak}}{R_s(i)} \mcomma\\
$}
\end{equation}
\begin{equation}
\resizebox{0.875\columnwidth}{!}{$
\charge{rx}^{l}(i, j) =
    \ln \left(\frac{V_{i}^{l}}{V_{i}^{l} + \current{peak} \cdot R_d(i, j)}\right) \cdot
    \frac{L \cdot V_{i}^{l}}{R_d(i, j)^2} \text{+}
    \frac{L \cdot \current{peak}}{R_d(i, j)} \mperiod
$}
\end{equation}
Before building our constraints, we need to define our \textbf{decision variable} $\ct{i, j}^{l}$, which denotes the number of transfer operations from the $i$-th cell to the $j$-th during the $l$-th idle period.
Notably, the amount of charge transmitted and received depends on the distance between the pairs of cells and their voltages.
The cell voltage is proportional to the \gls{soc} of cells.
In this paper, we assume that cell voltage remains constant during the balancing process as the difference in charge is negligible~\cite{Roy2019Optimal}.
We also assume that the balancing operations happen only when the pack is not utilized, i.e., during the idle periods.
We define $P_{i}$ as the set of ``compatible'' cells that can either receive from or transmit to the $i$-th cell, i.e., those within the maximum distance $d$.
We can start building our first constraint by computing each cell's total transmitted and received charges during each idle period.

The total transmitted charges from the $i$-th cell to all the other compatible cells in the $l$-th idle period is computed as:
\begin{equation}\label{eq:tx_l}
    \charge{T,i}^{l} = \sum_{j \in P_{i}} \left(\charge{tx}^{l}(i) \cdot \ct{i, j}^{l}\right) \mcomma
\end{equation}
while the charge the $i$-th cell receives is computed as follows:
\begin{equation}\label{eq:rx_l}
    \charge{R,i}^{l} = \sum_{j \in P_{i}} \left(\charge{rx}^{l}(i, j) \cdot \ct{j, i}^{l}\right) \mperiod
\end{equation}

The act of balancing inherently changes the charge level of the cells.
We can compute the charge level of the $i$-th cell after balancing at the $k$-th segment, as follows:
\begin{equation}\label{eq:bal_k}
    \overline{\charge{i}^{k}} =
        \charge{i}^{k} +
        \sum_{l \in \mathcal{L}_{k}}{\left(\charge{R,i}^{l} - \charge{T,i}^{l}\right)} \mperiod
\end{equation}

After a balancing operation, the charge levels must be sufficient for upcoming driving activities but always stay within safe ranges.
This means they must be \emph{higher} than the minimal usable capacity and \emph{lower} than the maximum nominal capacity.
As such, we can formulate our first balancing constraint as follows:
\begin{equation}\label{eq:const_min}
    \forall i,k, \;\;
        \charge{UC} \leq \overline{\charge{i}^{k}} \leq \maxcharge{i} \mperiod
\end{equation}
Here $k$ is unbounded (i.e., $\forall k$); however, in the objective function, $k$ will be set to the size of our \emph{knowledge window}.

As mentioned in \cref{subsec:balancing_architectures}, the time needed to perform a \emph{single transfer cycle}, namely $T_c$, depends on the specific pairs of cells selected for the operation.
We can compute the time it takes for the $i$-th cell to receive charges from the other compatible cells during the $l$-th idle period as follows:
\begin{equation}
    T_{tran, i}^{l} = \sum_{j \in P_{i}} \left( \ct{j, i}^{l} \cdot T_{c}(j,i) \right) \mcomma
\end{equation}
where $T_{c}(j, i)$ is the specific \emph{single transfer cycle} time between cells $j$ and $i$.
In general, the balancing time is given by the number of \emph{single transfer cycles} it takes each feasible pair of cells to perform charge equalization.
Here, we perform balancing during idle periods; thus, we must constrain the balancing time to complete within those periods.
This means that the balancing time during the $l$-th idle period must be lower than the length of the idle period $\Delta_l$.
We can write our second balancing constraint as follows:
\begin{equation}\label{eq:const_time}
    \forall l \in \mathcal{L}_{k}, \;\;
        \sum_{i} T_{tran, i}^{l} \leq \Delta_{l} \mperiod
\end{equation}

After defining these two constraints, we build our objective function to be minimized.
The battery's throughput up to the $k$-th segment, defined as the sum of absolute charged and discharged current, can be computed by treating the balancing as micro-charging or micro-discharging events as follows:
\begin{equation}
\resizebox{0.875\columnwidth}{!}{$
\ahthrp{i}^k
    = \int \left(\rate{i} \cdot |\current{}(t)| + \selfdischarge{i}\right)dt +
        \sum\limits_{l \in \mathcal{L}_{k}}{\left(\charge{T,i}^{l} + \charge{R,i}^{l}
        \right)} \mperiod
$}
\end{equation}
While the ratio of average current to maximum capacity up to the $k$-th segment is defined as:
\begin{equation}
    \crate{i}^k \;=\; \frac{\ahthrp{i}^k}{\maxcharge{i} \cdot T} \mcomma
\end{equation}
where $T$ is the cumulative usage time up to the $k$-th segment.

It would seem appropriate to minimize the capacity loss of the $i$-th cell by using \cref{eq:capacity_lost_general}.
However, since the aging model exhibits non-linear behavior, the objective function cannot be directly formulated in a linear form by using \cref{eq:capacity_lost_general}.
Instead, we have seen after extensive study that $\crate{i}$ is a monotonically increasing function of $\ahthrp{i}$, and the capacity loss is also a monotonically increasing function of $\ahthrp{i}$.
We can use $\ahthrp{i}$ in the formulation to maintain linearity instead of directly using \cref{eq:capacity_lost_general}.
Finally, we can build our optimization function to minimize the maximum degradation among the cells (i.e., wear leveling) for a given time window $w$.
When the number of mission segments for the next day is $w$, we can limit the range of values that $k$ can assume, and the \emph{wear leveling-aware} balancing objective can be formulated as follows:
\begin{equation}\label{eq:min_aware}
\begin{gathered}
    \mathbf{Min} \; \Big( max(\ahthrp{i}^k) \Big) \mcomma\\
    \mathbf{s.t.}  \; \forall i, t \leq k \leq t+w, \; Eqs.\ \eqref{eq:const_min}\ and\ \eqref{eq:const_time} \textit{ hold true,}
\end{gathered}
\end{equation}
where the current time is $t$, and $w$ is the size of our \emph{knowledge window}.
We write the optimization problem described above as an \gls{milp} formulation.
Its solution yields the number of charge transfer operations during idle periods $\ct{i, j}^{l}$, equalizing cells wear.

\section{Experimental setup and results}\label{sec:results}
This section compares our proposed \emph{wear leveling-aware} strategy against the state-of-the-art \emph{opportunistic} one in typical real-world scenarios.
Our simulation environment is written in Python, while we use CPLEX~\cite{CPLEX} to solve the \gls{milp} problem, with a timeout of \SI{5}{\second} for finding a solution.
If the optimizer takes more than \SI{5}{\second} to complete, we return the initial charge distribution as a result (i.e., no balancing).
This is necessary because the \emph{opportunistic} strategy tends to run for long periods, leading to simulations that last for hours.

We consider the peak current $\current{peak}$ for our battery pack to be \SI{12}{\ampere}, and the maximum distance between cells can exchange charge $d$ is 6.
We use the aging model shown in \cref{eq:capacity_lost_general} and set the temperature of the cells to \SI{22}{\celsius} for all experiments.
We interpolate parameters $a$ and $b$ for the \SI{22}{\celsius} temperature, based upon the values proposed in~\cite{Wang2014Degradation}, and obtain $a=0.00083$ and $b=0.3789$.
To model manufacturing process variation for our \SI{2.5}{\ampere\hour} cells, we generate their nominal capacities following a normal distribution with a mean of $100\%$ and a standard deviation of $4\%$~\cite{Dubarry2009Single}.
We have also randomly generated each cell's self-discharge currents, charge, and discharge rates following a normal distribution.
Specifically, we uniformly distribute charge and discharge rates in relatively small ranges with $\ratecharge{} \in [0.996, 1.00]$ and $\ratedischarge{} \in [1.00, 1.001]$, since \gls{lion} batteries are known to have high efficiency~\cite{Yang2018Study}.
On the other hand, since self-discharge currents are the leading causes of imbalance, we assume they follow a normal distribution with a mean of \SI{0.175}{\milli\ampere} and a standard deviation of \SI{0.1}{\milli\ampere}.
Even when the battery is not utilized, part of its charge is depleted because of the self-discharge phenomenon.
We set the minimum usable capacity $\charge{UC}$ as 20\% of the total capacity~\cite{Ovejas2019State}.

To compare the balancing algorithms extensively, we generated 50 scenarios with the following procedure.
First, we randomly generate a connected Watts–Strogatz small-world graph with 30 nodes, each connected with six nearest neighbors, and a 30\% probability of rewiring each edge (see~\cite{Watts1998Collective} for a detailed explanation).
Then, we randomly generate travel times and currents for each edge, select four nodes as charging stations, and generate probability distributions for the outgoing edges of each node.
We generate our missions by simulating thousands of random walks of length 10, starting from a node marked as a charging station.
That ensures we can charge the battery at least once during our missions.
We place an idle period with a randomly generated duration between segments.
\begin{table}[b]
    \vspace*{-2em}
    \caption{Minimum, maximum, and average peak memory usage and solve time for each balancing technique.}
    \label{tab:runtime_statistics}
    \vspace*{-1em}
    \centering
    \begin{tabular}{lr|r|r|r|r|r}
        \toprule
        \multirow{2}{*}{Algorithm} &
        \multicolumn{3}{c}{Peak Memory (KB)} &
        \multicolumn{3}{c}{Solve time (ms)}\\
        \cmidrule(lr){2-4}
        \cmidrule(lr){5-7}
        & Min & Avg& Max& Min & Avg & Max\\
        \midrule
        \multicolumn{1}{l|}{Opportunistic}
        &  2 & 28 & 561 & 33 & 2472 & 5489\\
        \multicolumn{1}{l|}{Wear leveling-aware}
        & 69 & 71 & 176 & 64 &   79 & 1268\\
        \bottomrule
    \end{tabular}
\end{table}

Solving an optimization problem can be both time-consuming and energy-intensive.
This could be problematic if the algorithm that solves it is meant to be used in an embedded setup.
As such, it is essential to assess both the time taken and the memory used by the different balancing techniques.
Before discussing the results, it is worth clarifying that the \emph{opportunistic} strategy runs at every idle period.
In contrast, the \emph{wear leveling-aware} one is executed only once at the start of the knowledge window.
The algorithm can work with knowledge windows ranging from knowing the next driving segment to knowing the entire day or several.
As such, the size of the knowledge window determines how many times the \emph{wear leveling-aware} optimization runs.
\cref{tab:runtime_statistics} reports the peak memory usage and the elapsed time required to solve each optimization problem, with \emph{wear leveling-aware} having a knowledge window of one day.
Both strategies have low average memory usage, making both viable solutions for an embedded system platform.
The \emph{Opportunistic} strategy takes considerably more time to find a solution than the \emph{wear leveling-aware} one.
The average solve time for \emph{opportunistic} balancing would have been even higher if we did not set a \SI{5}{\second} timeout inside the CPLEX solver.
The \emph{opportunistic} strategy has a broader range of performance metrics, while the \emph{wear leveling-aware} one has a narrower range.
These results show how the \emph{wear leveling-aware} strategy has lower variability with comparable if not better results than the \emph{opportunistic} one, making it a more reliable and compelling solution.

\begin{figure}[tbp]
    \centering
    \includegraphics[width=\columnwidth]{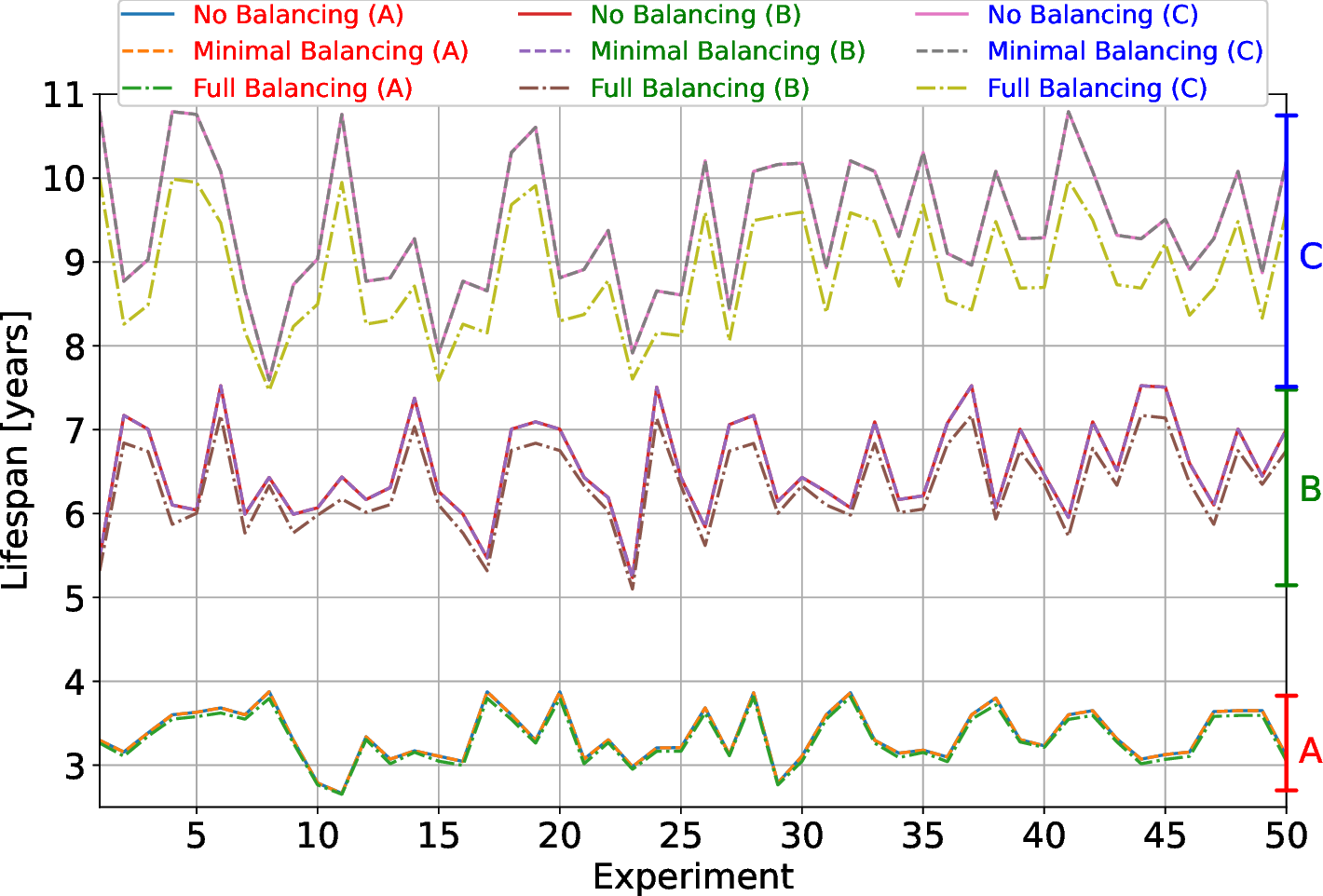}
    \caption{Comparing battery lifespan with \emph{none}, \emph{opportunistic} and \emph{wear leveling-aware} (WLA) balancing strategies.}
    \label{fig:exp_lifespan}
    \vspace*{-1.5em}
\end{figure}

Next, we evaluated the impact of each balancing strategy on the lifespan.
We are comparing the lifespan, in each mission, when the car is used every day (\textit{A}), once every two days (\textit{B}), and once every three days (\textit{C}).
This sums up to a total of 150 simulations and comparisons.
The comparison between the three strategies is depicted in \cref{fig:exp_lifespan}.
The \emph{wear leveling-aware} strategy has the same lifespan as \textbf{not doing balancing} in all scenarios.
When comparing the \emph{wear leveling-aware} against the \emph{opportunistic}, the improvements are, on average, one month with scenario \textit{A}, four and a half months with scenario \textit{B}, and around ten months with scenario \textit{C}.
Although all experiments are substantially different from each other, \emph{wear leveling-aware} can improve the lifespan in every one of them.
Let's compare the average number of balancing operations of \emph{wear leveling-aware} against the \emph{opportunistic}.
It is 7 against 842 in scenario \textit{A}, it is 3 against 420 in scenario \textit{B}, and it is 3 against 284 in scenario \textit{C}.
As expected, \emph{wear leveling-aware} drastically reduces the number of balancing operations.
Overall, \emph{wear leveling-aware} has better performances in terms of reduced aging, memory consumption, and solving time, making it a compelling cell balancing strategy.

\section{Concluding remarks}\label{sec:conclusion}
\glsresetall
This paper presents an active cell balancing strategy, viz., when and how much we balance, that optimally triggers balancing to minimize aging from balancing.
We compared this \emph{wear leveling-aware} strategy against a more intuitive \emph{opportunistic} strategy.
Results show that our approach ensures we can complete a planned driving mission while having a negligible impact on aging from balancing.
Although both strategies have a low impact on memory usage, \emph{opportunistic} balancing takes considerably more time and often fails to solve.
Furthermore, our approach is compatible with different cell aging models and balancing architectures.
Future investigations will study (1)~stocastic modeling of future driving patterns to replace the current oracle-like knowledge of them, (2)~include dynamic evolution of cell temperature based upon thermal modeling and simulation, and (3)~ testing the aging evolution with other optimization objectives.

\balance 
\bibliographystyle{IEEEtran}
\bibliography{bibliography}

\end{document}